\documentclass[epj]{svjour}
\usepackage{ulem}

\usepackage{graphics}

\usepackage{physics}
\usepackage{amsmath}
\usepackage{breqn}

\begin{document}
\title{Fermionic quantum cosmology as a framework for resolving type IV singularities

}
\subtitle{}
\author{Paweł Kucharski\inst{1}\thanks{\emph{Present address:} pawel.kucharski@phd.usz.edu.pl} \and Adam Balcerzak\inst{2}\thanks{\emph{Present address:} adam.balcerzak@usz.edu.pl}%
}                     
%
%
\institute{Doctoral School, University of Szczecin, Mickiewicza 18, 70-384 Szczecin, Poland \and Institute of Physics, University of Szczecin, Wielkopolska 15, 70-451 Szczecin, Poland}
\date{Received: 22 March 2025 / Revised version: 22 March 2025}
%
\abstract{
In this paper, we present an innovative approach to resolving type IV singularities in fermionic quantum cosmology. The Eisenhart-Duval lift procedure is employed to construct an extended minisuperspace metric, which allows for the formulation of the Dirac equation in minisuperspace. Through this approach, fermionic degrees of freedom are effectively incorporated into a homogeneous and isotropic cosmological model with a scalar field. By applying a kind of the Born-Oppenheimer approximation, solutions to the Dirac equation for an approximate potential characteristic of type IV singularities are obtained, expressed in terms of Tricomi confluent hypergeometric functions and associated Laguerre polynomials. The elimination of non-physical, divergent solutions results in a quantum regularization of the classical singularity. These results indicate the potential of fermionic models in quantum cosmology for mitigating the singularity problem.
} 
\maketitle
\section{Introduction}
\label{intro}

One of the greatest challenges of modern cosmology is to provide physically meaningful solutions to cosmological equations \cite{HH,Hawking,Kiefer0,Kiefer1}. Classical Einstein equations fail to do so, as they lead to singularities \cite{classicalsingularity}, which may indicate the need for a new theory of gravity.

In the case of Einstein’s equations, we can find solutions for a homogeneous and isotropic universe filled with a fluid characterized by pressure $p$ and density $\rho$. These solutions are known as the Einstein–Friedmann equations.
The singularities can be introduced to the equations in four different ways, all of them were shortly characterized below.
Type I singularity \cite{singularities1,sin1,singularities3} can be introduced for very large negative pressure $p<-\rho$.
This condition leads to a singularity in the future, which is known as the Big Rip. In this case, the scale factor $a(t)$ diverges in a finite time, leading to a singularity. During the expansion of the universe, the pressure approaches minus infinity, while the energy density tends toward infinity. In practice, this means that the expansion of the universe continues to accelerate, ultimately leading to the disintegration of galaxies, planetary systems, and even elementary particles.
A Sudden Future Singularity (type II singularity) \cite{singularities1,singularities3,sin2,sin2O1} represents an abrupt change in the dynamics of the universe while maintaining the finiteness of the scale factor and energy density. Geodesics can be extended through this singularity, even though the pressure and higher derivatives of the scale factor diverge.
A type III singularity \cite{chaplyginsingularity}, also known as a finite scale factor singularity, in contrast to a Type II singularity, involves the divergence of both pressure and energy density while the scale factor remains finite at a finite time. Another important difference is that this type of singularity typically does not allow for the extension of geodesics through the singular point.
The type IV singularity \cite{singularities1,singularities3,sin4,sin4WdW,sin4O1,sin4O2,sin4O3,sin4O4,sin4O5,sin4O6}, which is the focus of this publication, is the mildest among the types of singularities discussed. Geodesics can be smoothly extended through the singular point in finite time. The scale factor, energy density, and pressure all remain finite and regular; however, higher-order derivatives of the scale factor exhibit divergence. Notably, the type IV singularity can be avoided in standard quantum cosmology based on the Wheeler–DeWitt equation \cite{sin4WdW}.

In this publication, we make use of the method called Eisenhart–Duval lift \cite{EDlif1,EDlif2,EDlif3,EDlif4,EDlif5}, which serves as a bridge between field theory and differential geometry.
By increasing the number of dimensions, we can represent a dynamical system purely geometrically, where the system evolves along geodesics in a higher-dimensional space whose metric depends on the potential. The resulting metric, known as the extended minisuperspace metric, will be used to geometrize the Dirac equation.

Classical quantum cosmology - which is based on the Wheeler–DeWitt equation - faces several challenges \cite{Kiefer1}.
To address these issues, attempts have been made within the framework of supersymmetry \cite{moniz1,moniz2,moniz3,moniz4,moniz5}.
Another line of research involves taking the square root of the WdW equation in order to obtain an equation analogous to the Dirac equation \cite{Kan1,Kan2,Mallett}.
This latter approach represents the development of fermionic quantum cosmology provides an intriguing alternative to scalar quantum cosmology and offers several key advantages.
The Wheeler–DeWitt equation — being the analogue of the Klein–Gordon equation — is a second‑order equation with an indefinite norm, which gives rise to difficulties in the probabilistic interpretation of the Universe’s wave function \cite{Kiefer0,Kiefer1}.
By contrast, the fermionic approach removes this issue by introducing a Dirac equation, which is first‑order and guarantees the existence of a positive‑definite probability density for the Universe’s wave function \cite{Kim,Death,Yamazaki,Hojman}.
Furthermore, the requirement of full covariance in the extended minisuperspace fixes the operator ordering uniquely and allows for a consistent definition of the Hamiltonian operator’s square root \cite{factor_ordering1,factor_ordering2,factor_ordering3} which was a long‑standing ambiguities associated with operator ordering in the WdW equation.
The aim of this study is to investigate whether, in fermionic quantum cosmology constructed based on the Dirac equation and the metric for the extended minisuperspace, it is possible to obtain physically acceptable solutions. In particular, we are interested in whether such an approach allows for the removal of type IV singularities, which arise when the proposed potential is applied in classical cosmology.

Section 2 contains a recipe for introducing type IV singularity into the equations with specific form of potential. Section 3 presents the method for deriving the Dirac equation in the extended minisuperspace, as well as the procedure for obtaining the metric of this space. Section 4 contains the solution to the Dirac equation in the extended minisuperspace for the case of an approximate potential introducing a type IV singularity. Section 5 summarizes the results of the study.

\section{The appearance of a type IV singularity}
\label{singularityIV}

In order to consider a type IV singularity, it is necessary to adopt an appropriate cosmological model.
In this work, we analyze a perfect fluid in the form of a generalized Chaplygin gas, which - when used as a model of the Universe - can give rise to nearly all known types of singularities, including the type IV singularity \cite{chaplyginsingularity}. Originally proposed in the context of aerodynamics, it was later generalized and introduced as a candidate to describe both dark matter and dark energy \cite{chaplygineq1,chaplygin2,chaplygin3,chaplygin4,chaplygin5,chaplygineq2,chaplygin7}. One of its remarkable properties is its ability to smoothly interpolate between different cosmological epochs: in the early stages of the Universe's evolution, it behaves like pressureless dust, while in later times, it resembles a cosmological constant \cite{chaplygin5,chaplygineq2}.
The equation of state satisfied by the generalized Chaplygin gas is given by \cite{chaplygineq1,chaplygineq2}:
\begin{equation}
P = - \frac{A}{\rho^{\theta}},
\end{equation}
where $A$ and $\theta$ are constants.
A perfect fluid of this kind can be represented as a scalar field described by the equations:
\begin{equation}
\rho_\phi = \frac{1}{2} \dot{\phi}^2 + V(\phi), \quad p_\phi = \frac{1}{2} \dot{\phi}^2 - V(\phi),
\end{equation}
with a potential given by \cite{potential}:
\begin{equation}\label{eq:full_potential}
\begin{aligned}
V(\phi) = V_1 \Big[ &\sinh^{\frac{2}{1+\theta}} \left( \frac{\sqrt{3}}{2} \kappa |1+\theta|\, |\phi| \right) \\
&- \sinh^{-\frac{2\theta}{1+\theta}} \left( \frac{\sqrt{3}}{2} \kappa |1+\theta|\, |\phi| \right) \Big],
\end{aligned}
\end{equation}
\begin{figure}[tb]
  \centering
\resizebox{0.48\textwidth}{!}{\includegraphics{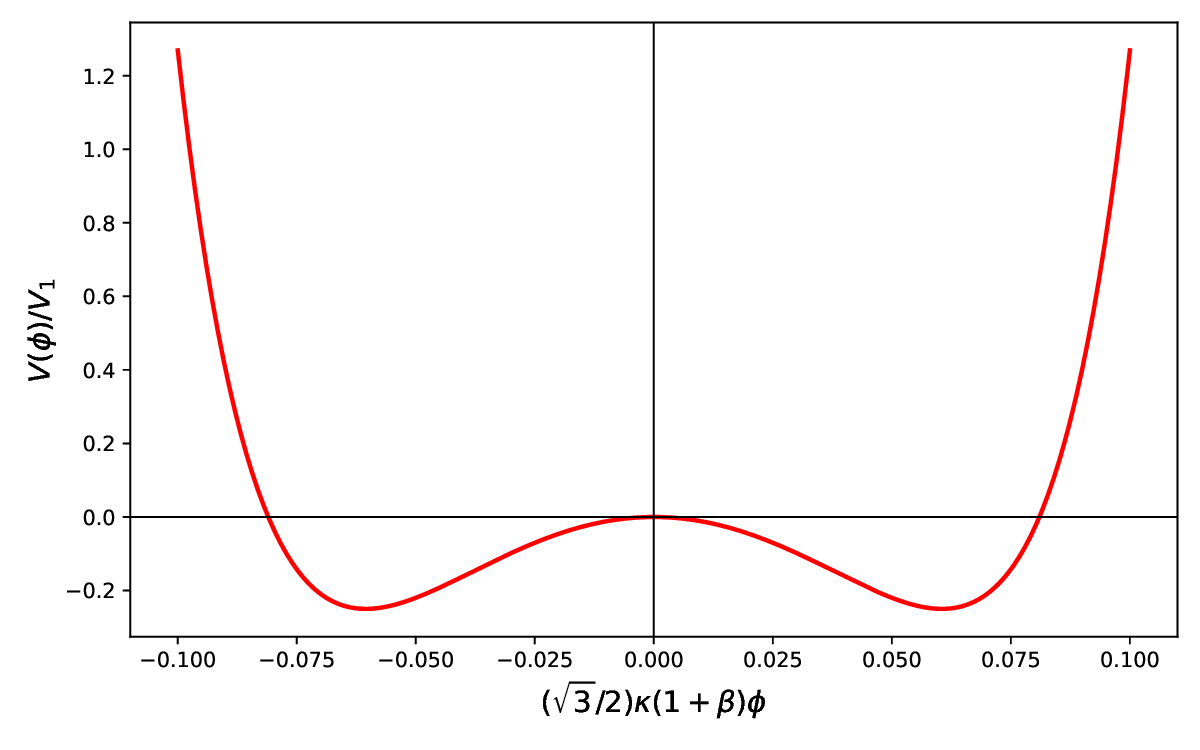}}
\caption{Double-well potential from Eq. \eqref{eq:full_potential} for $\theta = -\frac{1}{2}$.}
\label{fig:potential_plot}
\end{figure}
where $V_1$ is a constant, $\kappa^2 = 8\pi G$ where $G$ is the gravitational constant, and $\phi$ is a field. The potential $V(\phi)$ is a double-well potential, which is shown in Fig. \ref{fig:potential_plot}.
It can be shown that near the singular point the full potential given by the formula \eqref{eq:full_potential} can be approximated as \cite{potential}
\begin{equation}\label{eq:appx_potential}
  V(\phi) \simeq -V_1\left(\frac{\sqrt{3}}{2}\,\kappa\,|1+\theta|\,|\phi|\right)^{-\frac{2\theta}{1+\theta}}.
\end{equation}
In the singular case $\theta = -\frac{1}{2}$, this form of potential corresponds to an inverted harmonic oscillator.

\section{Dirac Equation in the Extended Minisuperspace}
\label{dirac_equation}

In this section, we derive the Dirac equation in an extended minisuperspace framework obtained by employing the Eisenhart-Duval lift on the standard minisuperspace metric for scalar field. The extended metric then helps to formulate the corresponding Dirac equation in fermionic framework.

\subsection{Extended Minisuperspace}
\label{extended_minisuperspace}

Consider a homogeneous and isotropic universe that contains a scalar field described by the following action \cite{Kan1}:
\begin{equation}
S = \int d^4x \sqrt{-g} \left( \frac{R}{2\kappa^2} - \frac{1}{2} g^{\mu\nu} \partial_{\mu} \phi \partial_{\nu} \phi - V(\phi) \right),
\end{equation}
where $R$ is the Ricci scalar, $\kappa^2 = 8\pi G$ with $G$ representing Newton's gravitational constant, $\phi$ denotes the real scalar field, and $V(\phi)$ is the scalar field potential.

Assuming homogeneity and isotropy, the metric takes the Friedmann-Lemaître-Robertson-Walker (FLRW) form:
\begin{equation}
ds^2 = -N^2(t) dt^2 + a^2(t) d\Omega^2_3,
\end{equation}
where $N(t)$ is the lapse function, $a(t)$ is the scale factor, and $d\Omega^2_3$ corresponds to the metric of a maximally symmetric three-dimensional space characterized by constant spatial curvature $K$, with Ricci curvature tensor $^{(3)}R_{ij}=2K g_{ij}$.

Using the above definitions, the resulting Lagrangian can be expressed as
\begin{equation}
L = -\frac{a\dot{a}^2}{2N} + \frac{a^3\dot{\phi}^2}{2N} - N U(a, \phi),
\end{equation}
where the dot represents differentiation with respect to time, and the effective potential $U(a, \phi)$ is defined as
\begin{equation}
U(a, \phi) = a^3 V(\phi) - \frac{K a}{2}.
\end{equation}

Introducing an additional degree of freedom through the Eisenhart-Duval lift procedure, we rewrite the extended minisuperspace Lagrangian as \cite{Kan1}:
\begin{equation}
\tilde{L} = -\frac{a\dot{a}^2}{2} + \frac{a^3\dot{\phi}^2}{2} + \frac{\dot{\chi}^2}{4U(a,\phi)} = \frac{1}{2} G_{MN} \dot{X}^M \dot{X}^N,
\end{equation}
with the extended coordinates defined as $X^M = (a, \phi, \chi)$, and the corresponding extended minisuperspace metric $G_{MN}$ given explicitly by
\begin{equation}\label{eq:extended_minisuperspace_metric}
G_{MN} = \begin{pmatrix}
-a & 0 & 0 \\
0 & a^3 & 0 \\
0 & 0 & \frac{1}{2U(a, \phi)}
\end{pmatrix}.
\end{equation}
This metric structure is essential for deriving the Dirac equation within the extended minisuperspace context.
\subsection{Dirac Equation}
\label{dirac_equation_subsection}
The general form of the Dirac equation in curved spacetime is expressed as \cite{Kan1}:
\begin{equation}
\bar{D} \Psi \equiv \hat{\gamma}^M D_M \Psi \equiv \gamma^A e^M_A D_M \Psi = 0,
\end{equation}
where $\Psi$ is the Dirac spinor, $\hat{\gamma}^M$ are the Dirac matrices in curved spacetime. The Dirac matrices in the chosen representation are given explicitly as:
${\gamma^1 = \sigma^1}$, ${\gamma^2 = i\sigma^2}$, and ${\gamma^3 = i\sigma^3}$, defined in terms of Pauli matrices $\sigma^1$, $\sigma^2$, and $\sigma^3$.
Covariant derivative $D_M$ is given by
\begin{equation}
D_M \equiv \partial_M - \frac{1}{8} \omega_{MAB} [\gamma^A,\gamma^B],
\end{equation}
where $\omega_{MAB}$ is the spin connection, which can be expressed in terms of the tetrad fields $e^M_A$ as:
\begin{equation}
\omega_{MAB} = \frac{1}{2} e^N_A \left( \partial_M e_{NB} -\Gamma^L_{MN} e_{LB}  \right) - (A \leftrightarrow B).
\end{equation}
Tetrad fields $e^M_A$ satisfy the relation $G_{MN} = \eta_{AB} e^A_M e^B_N $, where $\eta_{AB}$ is the Minkowski metric.

We utilize the metric derived for an extended minisuperspace obtained from the scalar field model, although the exact form of the general solution remains uncertain. Employing this metric (given explicitly in Eq.~\eqref{eq:extended_minisuperspace_metric}), the Dirac equation takes the following explicit form:
\begin{equation}
  \begin{aligned}
    &\left[ \sigma^1 \left(4U(\alpha,\phi)\frac{\partial}{\partial \alpha}
      + \frac{\partial U(\alpha,\phi)}{\partial \alpha}
      + 3U(\alpha,\phi)\right)\right.\\
    &\quad\left. + i\sigma^2 \left(4U(\alpha,\phi)\frac{\partial}{\partial \phi}
      + \frac{\partial U(\alpha,\phi)}{\partial \phi}\right)\right.\\
    &\quad\left. + i\sigma^3\,4\sqrt{2}\,e^{\frac{3\alpha}{2}}\,U(\alpha,\phi)^{3/2}\frac{\partial}{\partial \chi}
    \right]\Psi(\alpha,\phi, \chi) = 0,
  \end{aligned}
\end{equation}
where we have introduced the new variable $\alpha = \ln a$ for convenience.

The dimensionality of the extended minisuperspace metric can be reduced by decomposing the spinor wavefunction into two components as:
\begin{equation}
\Psi(\alpha, \phi, \chi) = \left( \begin{array}{c}
\psi_1(\alpha, \phi) \\
\psi_2(\alpha, \phi)
\end{array} \right) e^{ip\chi}.
\end{equation}

\section{Solution for Approximate Potential}
\label{approximate_potential}
To explicitly solve the Dirac equation derived in the previous section, one must specify the form of the potential $U(\alpha, \phi)$. We adopt the approximate potential introduced in Eq.~(\ref{eq:appx_potential}), choosing parameters ${K = 0}$ and ${\theta = -\frac{1}{2}}$, corresponding to a type IV singularity. Under these conditions, the potential simplifies to:
\begin{equation}\label{eq:approximate_potential}
U(\alpha, \phi) \propto e^{3\alpha}\phi^2.
\end{equation}

To solve this simplified Dirac equation, we apply the kind of Born-Oppenheimer approximation.
The B-O approximation originates from quantum chemistry, where it has proven highly successful \cite{BOapprox}. It was originally introduced to separate the motion of atomic nuclei in atoms and molecules from the motion of electrons. The key assumption is that the dynamics of the heavy atomic nuclei are significantly slower than those of the lighter electrons, so the electronic configuration adjusts almost instantaneously to changes in the nuclear positions. This idea has been transposed into the context of cosmology \cite{KieferBO}. In this setting, the separation concerns the "heavy" geometrical degrees of freedom, such as the scale factor, and the "light" degrees of freedom, such as matter fields. In our case, we apply the following decomposition:
\begin{equation}
\Psi(\alpha, \phi) = \rho(\alpha, \phi) \xi(\alpha),
\end{equation}
where $\rho(\alpha, \phi)$ depends on $\alpha$ only as a parameter, not as a dynamical variable.

Here we make use of the fact that having a solution $\rho$ to the second-order Klein–Gordon equation $\hat{D}\hat{D}\rho = 0$ allows us to obtain a solution to the first-order Dirac equation by acting Dirac operator $\hat{D}$ on solution $\rho$. Dirac operator $\hat{D}$ is constructed with potential of form given in Eq. \eqref{eq:approximate_potential}

\begin{equation}\label{eq:dirac_operator}
\begin{aligned}
\hat{D} = -2\sigma^1 \left( E(\alpha)-3 \right) + i \sigma^2 \left( 2 \frac{\partial}{\partial \phi} + \frac{1}{\phi}\right) \\+ \sigma^3 k_1 e^{3\alpha} \phi,
\end{aligned}
\end{equation}
where $k_1$ is a constant and $E(\alpha)$ is an energy coming from kind of Born-Oppenheimer approximation.

Thus, we find a solution for the bispinor $\rho$ as follows:
\begin{equation}
\begin{aligned}
\rho(\alpha, \beta) = \left( \begin{array}{c}
\rho_1(\alpha, \beta) \\
\rho_2(\alpha, \beta)
\end{array} \right),
\end{aligned}
\end{equation}
where spinor components $\rho_1$ and $\rho_2$ are defined as:

\begin{equation}
\begin{aligned}
  \rho_1 \propto C_1 U\left(\frac{P(\alpha)-4}{8},-1,\frac{\beta^2}{2}\right) \\+ C_2 L^{-2}_{\frac{4-P(\alpha)}{8}}\left(\frac{\beta^2}{2}\right) \\+
  D_1 U\left(\frac{Q(\alpha)-4}{8},-1,\frac{\beta^2}{2}\right) \\+ D_2 L^{-2}_{\frac{4-Q(\alpha)}{8}}\left(\frac{\beta^2}{2}\right),
\end{aligned}
\end{equation}

\begin{equation}
  \begin{aligned}
    \rho_2 \propto C_1 U\left(\frac{P(\alpha)-4}{8},-1,\frac{\beta^2}{2}\right) \\+ C_2 L^{-2}_{\frac{4-P(\alpha)}{8}}\left(\frac{\beta^2}{2}\right) \\-
    D_1 U\left(\frac{Q(\alpha)-4}{8},-1,\frac{\beta^2}{2}\right) \\+ D_2 L^{-2}_{\frac{4-Q(\alpha)}{8}}\left(\frac{\beta^2}{2}\right),
  \end{aligned}
  \end{equation}
where $U(a,b,x)$ denotes Tricomi confluent hypergeometric function, $L^m_n(x)$ represents associated Laguerre polynomials, and $C_1$, $C_2$, $D_1$, $D_2$ are integration constants. The functions $P(\alpha)$ and $Q(\alpha)$ depend explicitly on $\alpha$.
Variable $\beta$ is defined as $\beta = \sqrt{k_1 e^{-3\alpha}} \phi$.

The full solution of the Dirac equation is subsequently obtained by applying the operator defined in Eq.~\eqref{eq:dirac_operator} to the bispinor $\rho$, resulting in a solution composed of Tricomi confluent hypergeometric functions and associated Laguerre polynomials. The explicit form of the solution is given in the appendix A.

The component of the solution proportional to Tricomi confluent hypergeometric function is not physically acceptable, as it diverges at the singularity ${\beta = 0}$. Physical regularity thus demands setting the constant $C_1 = 0$ and $D_1 = 0$. After this modification, the resulting solution becomes regular and physically well-behaved as it contains only associated Laguerre polynomials, which vanish at the singularity point ${\beta = 0}$ showing that the classical type IV singularity doesn't appear in this theory.

\section{Conclusion}
\label{conclusion}
The analysis demonstrates that extending the configuration space via the Eisenhart-Duval lift procedure is an effective tool for resolving type IV singularities in fermionic quantum cosmology. Formulating the Dirac equation with use of the extended minisuperspace metric allowed for the inclusion of fermionic degrees of freedom, which resulted in physically acceptable solutions. The application of the kind of Born-Oppenheimer approximation, along with a detailed analysis of the approximate potential, enabled the elimination of divergent solutions, leading to a quantum regularization of the classical singularity. A comparable result was reported in \cite{sin4WdW}, where the avoidance of the type IV singularity within standard Wheeler–DeWitt cosmology was attributed to the appearance of a specific component of the wave function. In the present work, however, we have shown that the singularity is avoided unconditionally. The obtained results not only confirm the validity of employing fermionic models in quantum cosmology but also indicate directions for further research, including the development of more complex potential models.

The role of fermionic degrees of freedom in cosmology is further emphasized by the work of Tukhashvili and Steinhardt \cite{TukStein}. In their paper, they showed that spinor condensation - triggered when the Ricci curvature exceeds a critical threshold - can lead to a smooth, non-singular bounce. Although their approach is semi-classical, in contrast to our quantum treatment, both frameworks succeed in resolving cosmological singularities and highlight the key role played by fermionic degrees of freedom in this context.

At this point, we should discuss the kind of B-O approximation we have employed.
In the context of separating gravitational and matter degrees of freedom near a type IV singularity, it is important to note that this singularity is the mildest among finite-time cosmological singularities - the scale factor and its first and second derivatives remain regular, while the singularity appears only in the third derivative (“superacceleration”) \cite{singularities1,singularities3}. This behavior allows one to treat the scale factor as a slow variable compared to the more rapidly evolving scalar field. The necessity of employing a kind of Born-Oppenheimer approximation arises due to the presence of a mass term in the Wheeler-DeWitt equation, which prevents exact separability; in such cases, separation is only possible within an adiabatic framework \cite{KieferBO}. This approach has been successfully applied in the context of type IV singularities in standard quantum cosmology \cite{sin4WdW}. While alternative separation schemes can affect the definition of time in perturbative settings \cite{Kamenshchik1,Kamenshchik2}, such ambiguities do not arise in our homogeneous minisuperspace model.

Another aspect that must be addressed here is the analysis of the boundary conditions we imposed on the wave function of the Universe. The issue of boundary conditions is a fundamental aspect of any quantum theory, as not all mathematically valid solutions are physically acceptable, and additional criteria must be applied to select meaningful ones. In our case, the specific potential near the type IV singularity leads to a scattering-like situation, conceptually similar to that considered in \cite{GasperaniBC}, where the wave function of the universe reflects off a potential barrier in minisuperspace. We imposed a boundary condition that eliminates the divergent component associated with the Tricomi confluent hypergeometric function, thereby discarding nonphysical transmitted and reflected modes, much like the exclusion of non-normalizable solutions in standard quantum mechanics. This situation is somewhat analogous to certain cases in Dirac quantum mechanics, where formally suggestive but unphysical solutions imply particle pair creation. A more complete treatment, which would require taking into account the backreaction of perturbations, could potentially lead to a dynamical derivation of such boundary conditions; although it is not evident whether this would be achieved, it remains an interesting topic for further investigation.

It should be noted that the potential we used was approximated around the point
$\beta=0$, and this was done for a flat spacetime ($K=0$).

Finally, results we obtained in this paper not only confirm the validity of employing fermionic models in quantum cosmology but also open new perspectives for addressing more realistic and complex potential landscapes in the early Universe.

%
%
%
%
%

\appendix
\section{Appendix}
\label{sec:appendixA}
Full solution of the Dirac equation is given by:
\begin{equation}
    \begin{aligned}
    \hat{D}\rho(\alpha, \beta) = Z(\alpha, \beta)\left( \begin{array}{c}
    \gamma_1(\alpha, \beta) \\
    \gamma_2(\alpha, \beta)
    \end{array} \right),
    \end{aligned}
\end{equation}
where
\begin{equation}
  Z(\alpha,\beta) = \frac{e^{-3 \alpha -\frac{\beta ^2}{4}}}{2\sqrt{2}\,\beta^{2}k_1\,|\beta|^{2}}.
\end{equation}
We also define:
\begin{equation}
  \begin{aligned}
    F(\alpha) =2 E(\alpha) - 3.
  \end{aligned}
\end{equation}
Then, explicit form of $\gamma_1$ and $\gamma_2$ is given by:
\begin{dmath*}
\gamma_1 = \\
| \beta |  \left(-C_1 U\left(\frac{P(\alpha)-4}{2},-1,\frac{\beta ^2}{2}\right) \left(e^{\frac{3\alpha}{2}} \sqrt{k_1}+\beta  F(\alpha)\right)\\
+e^{\frac{3\alpha}{2}} \beta ^2 (-C_1) \sqrt{k_1} (P(\alpha)-4) U\left(\frac{P(\alpha)}{2}-1,0,\frac{\beta ^2}{2}\right)\\
-2 e^{\frac{3\alpha}{2}} \beta ^2 C_2 \sqrt{k_1} L_{-\frac{P(\alpha)}{8}-\frac{1}{2}}^{-1}\left(\frac{\beta ^2}{2}\right)\\
-e^{\frac{3\alpha}{2}} C_2 \sqrt{k_1} L_{\frac{4-P(\alpha)}{8}}^{-2}\left(\frac{\beta ^2}{2}\right)\\
+e^{\frac{3\alpha}{2}} \beta^2 D_1 \sqrt{k_1} Q(\alpha) U\left(\frac{Q(\alpha)}{2}-1,0,\frac{\beta ^2}{2}\right)\\
-4 e^{\frac{3\alpha}{2}} \beta ^2 D_1 \sqrt{k_1} U\left(\frac{Q(\alpha)}{2}-1,0,\frac{\beta ^2}{2}\right)\\
+2 e^{\frac{3\alpha}{2}} \beta ^2 D_1 \sqrt{k_1} U\left(\frac{Q(\alpha)-4}{2},-1,\frac{\beta ^2}{2}\right)\\
+e^{\frac{3\alpha}{2}} D_1 \sqrt{k_1} U\left(\frac{Q(\alpha)-4}{2},-1,\frac{\beta ^2}{2}\right)\\
+2 e^{\frac{3\alpha}{2}} \beta ^2 D_2 \sqrt{k_1} L_{\frac{4-Q(\alpha)}{8}}^{-2}\left(\frac{\beta ^2}{2}\right)\\
+2 e^{\frac{3\alpha}{2}} \beta ^2 D_2 \sqrt{k_1} L_{-\frac{Q(\alpha)}{8}-\frac{1}{2}}^{-1}\left(\frac{\beta ^2}{2}\right)\\
+e^{\frac{3\alpha}{2}} D_2 \sqrt{k_1} L_{\frac{4-Q(\alpha)}{8}}^{-2}\left(\frac{\beta ^2}{2}\right)\\
-\beta  C_2 F(\alpha) L_{\frac{4-P(\alpha)}{8}}^{-2}\left(\frac{\beta ^2}{2}\right)\\
+\beta  D_1 F(\alpha) U\left(\frac{Q(\alpha)-4}{2},-1,\frac{\beta ^2}{2}\right)\\
+\beta  D_2 F(\alpha) L_{\frac{4-Q(\alpha)}{8}}^{-2}\left(\frac{\beta ^2}{2}\right)\right)\\
-2 e^{\frac{3\alpha}{2}} \beta  \sqrt{k_1} |\beta|' \left(C_1 U\left(\frac{P(\alpha)-4}{2},-1,\frac{\beta^2}{2}\right)\\
+C_2 L_{\frac{4-P(\alpha)}{8}}^{-2}\left(\frac{\beta ^2}{2}\right)\\
-D_1 U\left(\frac{Q(\alpha)-4}{2},-1,\frac{\beta ^2}{2}\right)\\
-D_2 L_{\frac{4-Q(\alpha)}{8}}^{-2}\left(\frac{\beta ^2}{2}\right)\right)
\end{dmath*}
\begin{dmath*}
\gamma_2 =\\
\left(2 e^{\frac{3 \alpha}{2}} \beta  \sqrt{k_1} |\beta|' \left(C_1 U\left(\frac{P(\alpha)-4}{2},-1,\frac{\beta ^2}{2}\right)\\
+C_2 L_{\frac{4-P(\alpha)}{8}}^{-2}\left(\frac{\beta ^2}{2}\right)+D_1 U\left(\frac{Q(\alpha)-4}{2},-1,\frac{\beta ^2}{2}\right)\\
+D_2 L_{\frac{4-Q(\alpha)}{8}}^{-2}\left(\frac{\beta^2}{2}\right)\right)\\
+| \beta |  \left(U\left(\frac{P(\alpha)-4}{2},-1,\frac{\beta ^2}{2}\right) \left(e^{\frac{3 \alpha}{2}} C_1 \sqrt{k_1} -\beta  C_1 F(\alpha)\right)\\
+e^{\frac{3 \alpha}{2}} \beta^2 C_1 \sqrt{k_1} (P(\alpha)-4) U\left(\frac{P(\alpha)}{2}-1,0,\frac{\beta ^2}{2}\right)\\
+2 e^{\frac{3 \alpha}{2}} \beta ^2 C_2 \sqrt{k_1} L_{-\frac{P(\alpha)}{8}-\frac{1}{2}}^{-1}\left(\frac{\beta^2}{2}\right)\\
+e^{\frac{3 \alpha}{2}} C_2 \sqrt{k_1} L_{\frac{4-P(\alpha)}{8}}^{-2}\left(\frac{\beta ^2}{2}\right)\\
+e^{\frac{3 \alpha}{2}} \beta ^2 D_1 \sqrt{k_1} Q(\alpha) U\left(\frac{Q(\alpha)}{2}-1,0,\frac{\beta ^2}{2}\right)\\
-4 e^{\frac{3 \alpha}{2}} \beta ^2 D_1 \sqrt{k_1} U\left(\frac{Q(\alpha)}{2}-1,0,\frac{\beta ^2}{2}\right)\\
+2 e^{\frac{3 \alpha}{2}} \beta ^2 D_1 \sqrt{k_1} U\left(\frac{Q(\alpha)-4}{2},-1,\frac{\beta ^2}{2}\right)\\
+e^{\frac{3 \alpha}{2}} D_1 \sqrt{k_1} U\left(\frac{Q(\alpha)-4}{2},-1,\frac{\beta ^2}{2}\right)\\
+2 e^{\frac{3 \alpha}{2}} \beta ^2 D_2 \sqrt{k_1} L_{\frac{4-Q(\alpha)}{8}}^{-2}\left(\frac{\beta ^2}{2}\right)\\
+2 e^{\frac{3 \alpha}{2}} \beta ^2 D_2 \sqrt{k_1} L_{-\frac{Q(\alpha)}{8}-\frac{1}{2}}^{-1}\left(\frac{\beta ^2}{2}\right)\\
+e^{\frac{3 \alpha}{2}} D_2 \sqrt{k_1} L_{\frac{4-Q(\alpha)}{8}}^{-2}\left(\frac{\beta ^2}{2}\right)\\
-\beta  C_2 F(\alpha) L_{\frac{4-P(\alpha)}{8}}^{-2}\left(\frac{\beta ^2}{2}\right)\\
-\beta  D_1 F(\alpha) U\left(\frac{Q(\alpha)-4}{2},-1,\frac{\beta ^2}{2}\right)\\
-\beta  D_2 F(\alpha) L_{\frac{4-Q(\alpha)}{8}}^{-2}\left(\frac{\beta ^2}{2}\right)\right)\right).
\end{dmath*}

\end{document}